\documentclass[reprint,
superscriptaddress,
 amsmath,amssymb,
 aps,
 prl
]{revtex4-2}
\usepackage{graphicx}
\usepackage{dcolumn}
\usepackage{bm}
\usepackage{tensor}
\usepackage[dvipsnames]{xcolor}
\usepackage{soul}
\usepackage{svg}
\usepackage{subcaption}
\usepackage{tabularray}
\usepackage{physics}
\usepackage[utf8]{inputenc} 
\usepackage[T1]{fontenc}
\usepackage{dsfont}
\usepackage[symbol]{footmisc}
\usepackage[%
    colorlinks=true,
    pdfborder={0 0 0},
    citecolor=blue,
    linkcolor=blue
]{hyperref}
\usepackage{url}
\usepackage{soul}
\begin{document}
\setstcolor{red}
\preprint{APS/123-QED}
\title{Efficient $n$-qubit entangling operations via a superconducting quantum router}
\author{Xuntao Wu}
\affiliation{Pritzker School of Molecular Engineering, University of Chicago, Chicago IL 60637, USA}

\author{Haoxiong Yan}
\altaffiliation[Present address: ]{Applied Materials Inc, Santa Clara CA 95051, USA}
\affiliation{Pritzker School of Molecular Engineering, University of Chicago, Chicago IL 60637, USA}

\author{Gustav Andersson}
\affiliation{Pritzker School of Molecular Engineering, University of Chicago, Chicago IL 60637, USA}

\author{Alexander Anferov}
\altaffiliation[Present address: ]{Department of Physics, ETH Z\"urich, 8093 Zürich Switzerland}
\affiliation{Pritzker School of Molecular Engineering, University of Chicago, Chicago IL 60637, USA}


\author{Christopher R. Conner}
\affiliation{Pritzker School of Molecular Engineering, University of Chicago, Chicago IL 60637, USA}


\author{Yash J. Joshi}
\affiliation{Pritzker School of Molecular Engineering, University of Chicago, Chicago IL 60637, USA}

\author{Bayan Karimi}
\affiliation{Pritzker School of Molecular Engineering, University of Chicago, Chicago IL 60637, USA}

\author{Amber M. King}
\affiliation{Pritzker School of Molecular Engineering, University of Chicago, Chicago IL 60637, USA}

\author{Shiheng Li}
\affiliation{Department of Physics, University of Chicago, Chicago IL 60637, USA}

\author{Howard L. Malc}
\affiliation{Pritzker School of Molecular Engineering, University of Chicago, Chicago IL 60637, USA}

\author{Jacob M. Miller}
\affiliation{Department of Physics, University of Chicago, Chicago IL 60637, USA}

\author{Harsh Mishra}
\affiliation{Pritzker School of Molecular Engineering, University of Chicago, Chicago IL 60637, USA}


\author{Hong Qiao}
\affiliation{Pritzker School of Molecular Engineering, University of Chicago, Chicago IL 60637, USA}

\author{Minseok Ryu}
\affiliation{Pritzker School of Molecular Engineering, University of Chicago, Chicago IL 60637, USA}

\author{Jian Shi}
\affiliation{Department of Materials Science and Engineering, Rensselaer Polytechnic Institute, Troy NY 12180, USA}

\author{Andrew N. Cleland}
\email{anc@uchicago.edu}
\affiliation{Pritzker School of Molecular Engineering, University of Chicago, Chicago IL 60637, USA}

\date{\today}

\begin{abstract}
Quantum algorithms on near-term quantum processors are typically executed using shallow quantum circuits composed of one- and two-qubit gates. However, as circuit depth and gate number increase, gate imperfections and qubit decoherence begin to dominate, limiting algorithmic complexity. An alternative approach is to explore gates involving more than two qubits. In previous work (X. Wu et al., Physical Review X \textbf{14}, 041030 (2024)), we demonstrated a new superconducting qubit architecture with user-selectable two-qubit interactions via a reconfigurable router, used to connect pairs of qubits. Here, we leverage this novel architecture to realize programmable and efficient multi-qubit operations involving more than two qubits, resulting in faster preparation of multi-qubit entangled states with good fidelities. We also successfully apply model-free reinforcement learning to perform multi-qubit gates, including training a two-qubit controlled-Z gate as well as three-qubit controlled-SWAP and controlled-controlled-phase (Fredkin and Toffoli) gates. Higher $n$th-order gates may also be feasible, using our high-connectivity router design. This could provide a more efficient and higher-fidelity implementation of complex quantum algorithms and a more practical approach to quantum computation.
\end{abstract}

\maketitle
\textit{Introduction}---As quantum computers scale to larger sizes~\cite{Castelvecchi2023, Manetsch2025}, there is a growing need for more efficient parallel execution of quantum gates, particularly for implementing complex quantum algorithms~\cite{Layden2023, Arute2019, Morvan2024} and quantum error correction (QEC) codes~\cite{Bluvstein2023, Bravyi2024, Acharya2024, Wang2026}. Conventional approaches to realizing target quantum operations generally rely on decomposition into an elementary gate set~\cite{Barenco1995}, typically of single- and two-qubit gates, which often result in lengthy and complex gate sequences. These overheads are further compounded by error-imposed limits on the achievable quantum volumes of quantum processors~\cite{Cross2019,GonzlezGarca2022}. More efficient processing can be achieved by entangling more qubits in parallel, including synthesizing native multi-qubit gates~\cite{Song2017,Feng2020, Roy2020,Gu2021,Kim2022,Feng2022, Glaser2023, Christensen2023, Warren2023, Hu2023, Itoko2024, Huai2024, Pratapsi2024, Liu2025, Zhao2025, Tasler2025}, providing more connectivity than in typical qubit layouts, as well as engineering analog system dynamics~\cite{Neeley2010, Kang2016,Song2019,Liu2020, Babukhin2020, GonzalezRaya2021, Yarkoni2022, Menke2022, Nguyen2024, Liang2024, Rosen2024, Zhang2024, Li2025, Zhao2025tunint, Yan2025, Huang2026}. Multi-qubit operations that extend beyond the conventional single- and two-qubit universal gate set can offer higher parallelism and reduced gate counts, while enabling more efficient parity-check operations and thereby faster QEC cycles~\cite{Huai2024, Tasler2025}.

Superconducting qubits are often designed in a planar layout, which supports nearest-neighbor connectivity but limits parallelism and in turn excludes classes of more complex quantum operations. Alongside the development of distributed and modular quantum computing architectures~\cite{Leung2019, Magnard2020, Zhong2021, Niu2023, Storz2023, Zhou2023, Field2024, Wu2024, Mollenhauer2025, Song2025}, quantum processors with higher connectivity and programmable couplings~\cite{Song2019, Zhou2023, Wu2024} provide compelling alternatives to strictly nearest-neighbor qubit interactions. However, as the number of interacting qubits increases, maintaining precise and robust control over increasingly complex higher-dimensional physical parameters becomes a challenge. This can be addressed in part by employing e.g. reinforcement learning (RL) techniques~\cite{Fsel2018, Niu2019, Zhang2019, Metz2023}, which offer universal control capabilities, support concurrent feedback-based training, and are robust against environmental noise. RL has been successfully implemented in a number of experimental quantum systems~\cite{Baum2021, Sivak2022, Sivak2023,Ding2023, Reuer2023, Wright2023, Reinschmidt2024, Chatterjee2024, Almanakly2025}, with applications to gate optimization~\cite{Baum2021, Ding2023, Wright2023}, state preparation~\cite{Sivak2022, Reuer2023}, transfer~\cite{Almanakly2025} and readout~\cite{Chatterjee2024}, error correction~\cite{Sivak2023}, and hardware control~\cite{Reinschmidt2024}. Here we describe integration of a particular RL algorithm to perform algorithms on our high-connectivity quantum processor ~\cite{Mohseni2024, Bausch2024, Alexeev2025}.

We have previously demonstrated a modular quantum processor with a central router~\cite{Wu2024}, supporting reconfigurable interactions between arbitrary qubit pairs. We demonstrated generation of high-dimensional entangled states across various qubits via sequential two-qubit interactions. The all-to-all connectivity of this design however supports the direct synthesis of multi-qubit operations beyond two-qubit gates, by simultaneously and dynamically turning on connections between more than two qubits. Here we demonstrate simultaneously coupling three and four qubits to one another, enabling faster and more efficient generation of multi-qubit entangled states, compared to sequential methods. We also train an RL agent to perform quantum gates, including a two-qubit controlled-Z (CZ) gate, with improved performance compared to a manually tuned-up gate, as well as three-qubit gates, implementing a controlled-SWAP (CSWAP) gate (also termed a quantum Fredkin gate~\cite{Patel2016}), as well as a controlled-controlled-phase (CCPHASE) gate~\cite{Song2017,Roy2020,Glaser2023}, a key building block for a Toffoli gate. We also explore analog quantum dynamics using a model Hamiltonian~\cite{Liu2020}.

\begin{figure}[tb]
    \centering
    \includegraphics{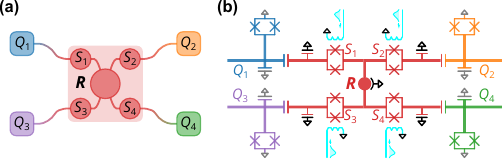}
    \caption{Experimental setup~\cite{Wu2024}. (a) Schematic and (b) circuit diagram of the experimental system, with four qubits coupled via a central router, whose active elements comprise four tunable switches.} \label{fig:fig1}
\end{figure}

\textit{Physical platform}---As illustrated in Fig.~\ref{fig:fig1}, our quantum processor~\cite{Wu2024} comprises two qubit-bearing sapphire dies (daughterboards), flip-chip assembled on a common larger sapphire die (the motherboard), which hosts the router as well as control and readout circuitry. By dynamically modulating the switches using local magnetic flux pulses, we can mediate many-to-many qubit couplings in a user-selected fashion, enabling multi-qubit interactions. Single-qubit gates as well as qubit readout are performed with the router coupling turned off. More details are provided in the Supplemental Material~\cite{Supplement}, and in Ref.~\cite{Wu2024}. 

\begin{figure}[tb]
    \centering
    \includegraphics{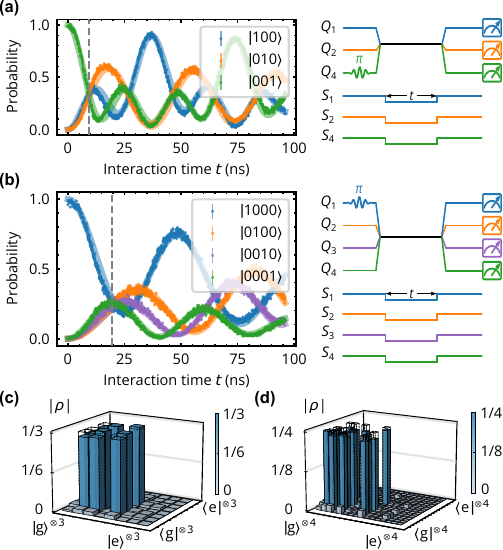}
    \caption{Generation of $W$ states via three- and four-qubit interactions. Panels (a) and (b) show the population evolution of a single excitation in simultaneous (a) three- and (b) four-qubit interaction, respectively, with gray dashed lines indicating the interaction time ($9.3~\mathrm{ns}$ and $19.6~\mathrm{ns}$) for generating the three- and four-qubit $W$ states, whose density matrices appear in panels (c) and (d). Light solid traces represent simulations adjusted to fit the experimental data (solid dots), using the Lindblad equation in a single-excitation truncated Hilbert subspace. Pulse sequences are shown on the right, with the simultaneous switch-controlling flux pulses during the interaction periods. (c) Three-qubit $W$ state density matrix with a fidelity $\mathcal{F}=\mathrm{tr}\left( \rho_{\mathrm{ideal}}\rho \right)=97.72(3)\%$. (d) Four-qubit $W$ state density matrix with a fidelity $\mathcal{F}=\mathrm{tr}\left(  \rho_{\mathrm{ideal}}\rho \right)=94.70(4)\%$.} \label{fig:fig2}
\end{figure}

\textit{Efficient entanglement generation}---We generate via a single-step process three- and four-dimensional entangled states, by coupling multiple qubits in parallel via the router. As an example, we describe the process for three- and four-qubit $W$ states. In previous work~\cite{Wu2024}, we used a sequential partial-swap method~\cite{Haffner2005} to generate both three-qubit and four-qubit $W$ states, achieving fidelities $\mathcal{F}=\mathrm{tr}\left( \rho_{\mathrm{ideal}} \rho \right)=93.87(7)\%$ and $88.87(8)\%$, with a total interaction time of $44.0~\mathrm{ns}$ and $56.6~\mathrm{ns}$, respectively. However, this method involves sequentially connecting different pairs of qubits, requiring the challenging minimization of signal interference and crosstalk. This also requires the high-fidelity tune-up of multiple partial swaps, which does not scale well with increasing qubit numbers.

Such states can instead be generated using a one-step pulse sequence in the spirit of Ref.~\cite{Neeley2010}, where a single excitation in one of the qubits is distributed across all coupled qubits until the desired final state is generated. As shown in Fig.~\ref{fig:fig2}(a) and (b), we use this method to generate three-qubit (four-qubit) $W$ states by first exciting $Q_4$ ($Q_1$), then simultaneously turning on the coupling to the other qubits. We leave the coupling on for the optimal interaction time ($9.3~\mathrm{ns}$ and $19.6~\mathrm{ns}$, respectively), so that the single excitation is evenly distributed among all the qubits, as indicated by the gray dashed lines in Fig.~\ref{fig:fig2}(a) and (b). After generating the $W$ states, we turn off all the switches and perform state tomography by applying scrambling single-qubit rotations $\in\left\{ I, X/2, Y/2 \right\}^{n}$ followed by multiplexed qubit readout. We find that we can obtain $W$ states with higher fidelities ($\mathcal{F}=\mathrm{tr}\left( \rho_{\mathrm{ideal}}\rho \right)=97.72(3)\%$ for three qubits in Fig.~\ref{fig:fig2}(c) and $94.70(4)\%$ for four qubits in Fig.~\ref{fig:fig2}(d)), with shorter preparation times, compared to the sequential method involving only single- and two-qubit gates. It is worth noting that even though the pulse sequences in Fig.~\ref{fig:fig2} indicate a resonance condition for the qubits, this is not technically necessary as long as the desired population distributions can still be realized.

In addition to preparing $W$ states, we generate Greenberger–Horne–Zeilinger (GHZ) states using a similar all-to-all interaction~\cite{Song2019}. Unlike the $W$ states, the protocol for GHZ states generally necessitates reaching specific resonance conditions. Furthermore, the inter-qubit coupling strengths should all be equal, ensuring the interacting qubits are effectively indistinguishable, thus realizing an equivalent one-axis twisting Hamiltonian. Meeting both criteria simultaneously poses a significant challenge. We therefore use reinforcement learning (RL) techniques, discussed below, to facilitate this process. We are able to generate e.g. a three-qubit GHZ state with a fidelity of $\mathcal{F}=\mathrm{tr}\left( \rho_{\mathrm{ideal}}\rho \right)=86.58(8)\%$. Details are provided in the Supplemental Material~\cite{Supplement}.

\textit{Synthesizing multi-qubit gates with RL}---We now turn from efficient generation of multi-qubit states to the construction of multi-qubit gates, which are pivotal for many quantum circuits. The entangling gates demonstrated here are all population-preserving, i.e. the number of photons in the system ideally remains constant during circuit evolution. Multi-qubit gates of this form can in principle be synthesized by only manipulating the flux in the system. However, manually calibrating multi-qubit collective dynamics by pairwise two-body interactions incurs a quadratic measurement overhead~\cite{Supplement}, which can be facilitated by quantum optimal control (QOC) algorithms~\cite{Ansel2024}. Modern QOC methods such as GRAPE~\cite{Khaneja2005}, Nelder-Mead~\cite{Kelly2014}, and Bayesian optimization~\cite{Granade2012} have been quite successful. However, in larger systems with more control parameters as well as fluctuating environments, more robust techniques such as RL can yield better performance~\cite{Sivak2022}, as these scale more easily to higher dimensions as well as can better manage relatively noisy environments.

\begin{figure}[tb]
    \centering
    \includegraphics{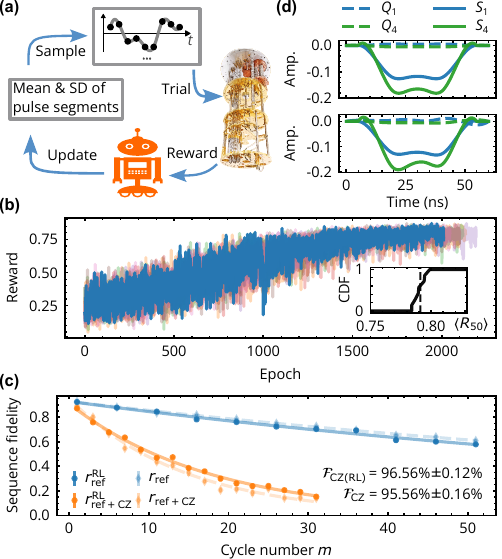}
    \caption{RL setup and CZ gate training. (a) RL optimization process. Pulse segments (black dots) are randomly generated using the current values for the mean and standard deviation of the pulse training parameters, which are then interpolated (gray line) and fed into the quantum system for evolution. The measurement results are used to evaluate a reward, which is passed to the agent to further optimize the training parameters. (b) Reward evolution during the training process. The reward definition as well as other details of the algorithm are given in the Supplemental Material~\cite{Supplement}. Traces with different colors are different realizations of the same process, all showing similar trends. Inset shows the CDF of the average of last $50$ rewards $\left\langle R_{50} \right\rangle$, indicating consistent convergence. (c) Cross-entropy benchmarking (XEB) of the RL-trained CZ gate (circle dots and solid lines, corresponding to the training indicated by the blue trace in panel (b)), yielding a benchmarked CZ fidelity of $96.56\%\pm0.12\%$, to be compared with the human-tuned CZ gate with flat-top flux pulses (diamond dots and dashed lines) featuring a slightly lower benchmarked fidelity of $95.56\%\pm0.15\%$. (d) Pulse shapes before (upper panel) and after (lower panel) the training process.} \label{fig:fig3}
\end{figure}

Our RL setup is illustrated in Fig.~\ref{fig:fig3}(a), where we employ the commonly-used proximal policy optimization (PPO) algorithm~\cite{Schulman2017}, due to its tailored training stability with a clipped objective function. In order to discretize the control problem, we parameterize each flux pulse with a piecewise constant function, with the mean and standard deviation (SD) values of each pulse segment as training parameters. At the beginning of each epoch (one loop in Fig.~\ref{fig:fig3}(a)), we sample a batch of pulse segments for all flux controls using the current mean and SD values. We then smooth the pulse shapes by interpolating based on the sampled pulse segments, before evolving the quantum system with the smoothed pulses. The reward is calculated from the measurement results, with the reward value used by the agent to update the training parameters. This process is repeated until convergence. We provide a detailed pseudocode in the Supplemental Material~\cite{Supplement}.

We first implement this method to tune up our two-qubit CZ gate. To eliminate the effect of state preparation and measurement (SPAM) errors on the optimization process, we introduce a method called optimized cross-entropy benchmarking for immediate tuneup (OXEBIT) which integrates the RL training with cross-entropy benchmarking (XEB)~\cite{Supplement}, in the same spirit as the ORBIT method~\cite{Kelly2014}. The reward is defined as $\mathrm{min}\left\{\sum_xp_{\mathrm{ideal}}(x)p_{\mathrm{exp}}(x)/\sum_xp_{\mathrm{ideal}}^{2}(x)\right\}$, where $p_{\mathrm{ideal}}(x)$ ($p_{\mathrm{exp}}(x)$) represents the ideal (experimental) probability of measuring the basis state $x$ after a circuit depth $m$~\cite{Wu2024}. The \emph{min} function runs over all $k$ random circuits, where we choose $m=3$ and $k=10$. We run the optimization algorithm on the $Q_1$-$Q_4$ CZ gate, with the evolution of the reward shown in Fig.~\ref{fig:fig3}(b); different colors represent different realizations of the training process with the same initial seed. We also display as inset to panel b the cumulative distribution function (CDF) of the average of the last $50$ rewards, denoted as $\left\langle R_{50} \right\rangle$. From the distribution, we have $\langle R_{50} \rangle = 0.791(5)$ where the uncertainty represents one standard deviation. This shows a consistent convergence of the training process. Note there is a sudden drop in the reward for the blue data at $\sim 1000$ epochs, likely from jumps in flux bias values, but the reward recovers within a few tens of epochs. The training takes about $2000$ epochs to converge, which corresponds to a total wall time of about $19$ hr, see Supplemental Material~\cite{Supplement}. Following the training, we perform XEB on the optimized pulses, and observe a $\sim 25\%$ reduction in the benchmarked infidelity compared to the manually tuned-up CZ gate, shown in Fig.~\ref{fig:fig3}(c). Moreover, as we can set an ideal unitary matrix as a training target, the RL-optimized CZ gate is free of accompanying single-qubit phases, which is more convenient for the quantum circuit design. In Fig.~\ref{fig:fig3}(d) we display the control pulse shapes before and after training.

We also use RL to directly synthesize two different three-qubit gates, namely the controlled-SWAP (CSWAP) and the controlled-controlled-phase (CCPHASE) gates. Multi-qubit gates outside the conventional single- and two-qubit universal gate set offer higher parallelism and possibly lower gate count, and support more efficient parity-check operations~\cite{Huai2024, Tasler2025}, yielding faster quantum error correction (QEC) cycles. As both the CSWAP and CCPHASE gates are population-conserving, these gates can be synthesized by manipulating just the qubit and router fluxes, thereby simultaneously rearranging the qubit excitations and accumulating the desired qubit phases.

\begin{figure}[tb]
    \centering
    \includegraphics{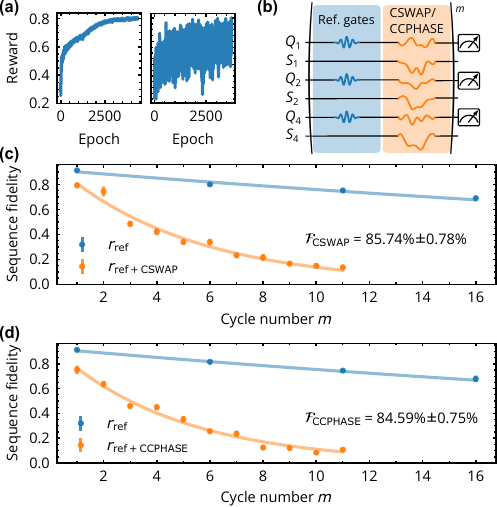}
    \caption{Three-qubit gate synthesis with RL. (a) Evolution of the reward for the CSWAP (left) and CCPHASE (right) gates. (b) XEB pulse sequence. Blue (orange) shaded area indicates the single-qubit reference gates (exemplary machine-learned three-qubit interleaved gates). XEB results for the (c) CSWAP and (d) CCPHASE gates, with fidelities $\mathcal{F}_{\mathrm{CSWAP}}=85.74\%\pm0.78\%$ and $\mathcal{F}_{\mathrm{CCPHASE}}=84.59\%\pm0.75\%$.} \label{fig:fig4}
\end{figure}

In the CSWAP gate, the states of the two target qubits swap according to the state of the control qubit, which can be expressed as $\mathcal{U}_{\mathrm{CSWAP}}=\dyad{0}{0}\otimes\mathds{1}+\dyad{1}{1}\otimes\left( \dyad{00}{00}+\dyad{10}{01}+\dyad{01}{10}+\dyad{11}{11} \right)$. As this is not a longitudinal gate, in addition to the flux-tuning-induced single-qubit phases, there are parasitic phases acquired due to mismatch of the qubit rotating frames~\cite{Sung2021, Wu2024}, as appear for a two-qubit iSWAP gate. This results in a more complicated randomized benchmarking process, which in turn affects the RL optimization procedure that relies on this benchmarking protocol. Here we instead choose to optimize using only the populations (we equivalently optimize for the initial computational basis states, namely $\lbrace \ket{000}, \ket{001}, \dots, \ket{111}\rbrace$, then determine the accompanying phases by qubit tomography. The reward is defined as the minimum fidelity of all target states. The evolution of the training for the reward is shown on the left side of Fig.~\ref{fig:fig4}(a). After training, we perform XEB by applying the pulse sequence in Fig.~\ref{fig:fig4}(b), and obtain a benchmarked fidelity of $\mathcal{F}_{\mathrm{CSWAP}}=85.74\%\pm0.78\%$, shown in Fig.~\ref{fig:fig4}(c). Following Ref.~\cite{Abad2022}, we estimate that decoherence contributes $8.14\%$~\cite{Supplement} to the total gate error ($14.26\%$), with the remaining infidelity attributed to enhanced decoherence during gate operation, leakage error, and other operational imperfections. Notably, decomposing the CSWAP gate into one- and two-qubit gates requires eight CNOT gates~\cite{MQCruz2024}, which would yield a synthesized $\mathcal{F}_{\mathrm{CSWAP}}$ fidelity below $75\%$, given the CZ fidelity we obtained in Fig.~\ref{fig:fig3}.

For the CCPHASE gate, there is a conditional $\pi$ phase accumulated on the target qubit conditioned on the states of the two control qubits. Given the specific frequency arrangement of our setup~\cite{Supplement}, the CCPHASE gate we optimize is given by $\mathcal{U}_{\mathrm{CCPHASE}}=\dyad{10}{10}\otimes\left( -\sigma_z \right) + \left( \mathds{1} - \dyad{10}{10} \right)\otimes\mathds{1}$. As the CCPHASE gate is also a longitudinal gate, similar to the CZ gate, we again train the gate using an integrated XEB process, with the reward evolution shown on the right of Fig.~\ref{fig:fig4}(a). After training, using the XEB sequence in Fig.~\ref{fig:fig4}(b), we obtain a benchmarked fidelity of $84.59\%\pm0.75\%$. Again, we estimate a decoherence contribution of $10.17\%$ out of $15.41\%$ gate infidelity~\cite{Supplement}, indicating that decoherence is the dominant factor limiting gate performance.

To the best of our knowledge, this is the first time an RL agent has been used for the synthesis of native three-qubit gates, illustrating the potential for RL in developing more efficient large-scale quantum computers. More details related to our optimization of these three-qubit gates are provided in the Supplemental Material~\cite{Supplement}.

\begin{figure}[tb]
    \centering
    \includegraphics{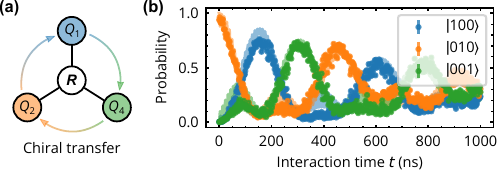}
    \caption{Spin chirality demonstration. (a) Schematic population transfer, where the qubit population flows follow a circular path $Q_2 \rightarrow Q_1 \rightarrow Q_4 \rightarrow Q_2 \rightarrow \cdots$. This pattern is performed experimentally as shown in (b), where points are experimental data, and solid lines are from simulations.} \label{fig:fig5}
\end{figure}

\textit{Analog quantum dynamics}---A natural application for a coupled multi-qubit system is the simulation of analog quantum dynamics under specific Hamiltonian models. Here, we take the spin chirality model~\cite{Liu2020} as an example to showcase the flexibility of the router design for synthesizing quantum dynamics. The goal is to couple three qubits with equal coupling strengths, then periodically modulate the qubit frequencies with appropriate phases. The resulting Hamiltonian in the rotating frame, using Floquet analysis, becomes proportional to the spin chirality interaction $\bm{\chi}=\bm{\sigma_1}\cdot\left( \bm{\sigma_2}\times\bm{\sigma_3} \right)$, where $\bm{\sigma_j}=\left( \sigma_j^x, \sigma_j^y, \sigma_j^z \right)$ is the Pauli vector~\cite{Liu2020}. This interaction results in a circular flow of excitations in the system, the direction determined by the externally-controlled phase modulation. During the interaction, the frequency of each qubit is modulated as $\omega_j(t)=\omega_0 + \Delta\cos\left( \nu t + \phi_j \right)$, with phases $\phi_{1,2,4}=2\pi/3,0,4\pi/3$. The population in the system then flows clockwise, as displayed in Fig.~\ref{fig:fig5}(a). The experimental data is shown in Fig.~\ref{fig:fig5}(b), where the decay in overall probability is limited by qubit dephasing and anharmonicity. We can also exploit the qubit frequency modulation to implement a tunable interaction between different pairs of qubits; see Supplemental Material~\cite{Supplement}.

\textit{Conclusion and outlook}---We have implemented efficient two-, three- and four-qubit operations on a router-centric modular superconducting quantum processor~\cite{Wu2024}, enabling the fast preparation of multi-qubit entangled states with good fidelities, as well as synthesis of native three-qubit gates (CSWAP and CCPHASE), achieved by training a reinforcement learning agent. We also demonstrate the flexibility of the router circuit by simulating analog quantum dynamics with precise hardware control. These results underscore the utility of high-connectivity architectures for exploring native multi-qubit interactions as well as engineering complex many-body quantum dynamics in a scalable fashion, paving the way for more resource-efficient implementations of advanced quantum algorithms and quantum error correction codes~\cite{Bravyi2024, Wang2026}.

\textit{Acknowledgments}---This work is supported by the Army Research Office and Laboratory for Physical Sciences (ARO grant W911NF2310077), the Air Force Office of Scientific Research (AFOSR grant FA9550-20-1-0270 and MURI grant FA9550-23-1-0338), DARPA DSO (grant HR0011-24-9-0364), and in part by UChicago's MRSEC (NSF award DMR-2011854), by the NSF QLCI for HQAN (NSF award 2016136), by the Simons Foundation (award 5099) and a 2024 Department of Defense Vannevar Bush Faculty Fellowship (ONR N000142512032). Results are in part based on work supported by the U.S. Department of Energy Office of Science National Quantum Information Science Research Center Q-NEXT. We made use of the Pritzker Nanofabrication Facility, which receives support from SHyNE, a node of the National Science Foundation's National Nanotechnology Coordinated Infrastructure (NSF Grant No. NNCI ECCS-2025633). The authors declare no competing financial interests. Correspondence and requests for materials should be addressed to A. N. Cleland (anc@uchicago.edu).
\bibliography{ref}

\end{document}


\setstcolor{red}
\title{Supplemental Material: Efficient $n$-qubit entangling operations via a superconducting quantum router}
\author{Xuntao Wu}
\affiliation{Pritzker School of Molecular Engineering, University of Chicago, Chicago, IL 60637, USA}

\author{Haoxiong Yan}
\altaffiliation[Present address: ]{Applied Materials Inc, Santa Clara CA 95051, USA}
\affiliation{Pritzker School of Molecular Engineering, University of Chicago, Chicago, IL 60637, USA}

\author{Gustav Andersson}
\affiliation{Pritzker School of Molecular Engineering, University of Chicago, Chicago, IL 60637, USA}

\author{Alexander Anferov}
\altaffiliation[Present address: ]{Department of Physics, ETH Z\"urich, 8093 Zürich, Switzerland}
\affiliation{Pritzker School of Molecular Engineering, University of Chicago, Chicago, IL 60637, USA}


\author{Christopher R. Conner}
\affiliation{Pritzker School of Molecular Engineering, University of Chicago, Chicago, IL 60637, USA}


\author{Yash J. Joshi}
\affiliation{Pritzker School of Molecular Engineering, University of Chicago, Chicago, IL 60637, USA}

\author{Bayan Karimi}
\affiliation{Pritzker School of Molecular Engineering, University of Chicago, Chicago, IL 60637, USA}

\author{Amber M. King}
\affiliation{Pritzker School of Molecular Engineering, University of Chicago, Chicago, IL 60637, USA}

\author{Shiheng Li}
\affiliation{Department of Physics, University of Chicago, Chicago, IL 60637, USA}

\author{Howard L. Malc}
\affiliation{Pritzker School of Molecular Engineering, University of Chicago, Chicago, IL 60637, USA}

\author{Jacob M. Miller}
\affiliation{Department of Physics, University of Chicago, Chicago, IL 60637, USA}

\author{Harsh Mishra}
\affiliation{Pritzker School of Molecular Engineering, University of Chicago, Chicago, IL 60637, USA}


\author{Hong Qiao}
\affiliation{Pritzker School of Molecular Engineering, University of Chicago, Chicago, IL 60637, USA}

\author{Minseok Ryu}
\affiliation{Pritzker School of Molecular Engineering, University of Chicago, Chicago, IL 60637, USA}

\author{Jian Shi}
\affiliation{Department of Materials Science and Engineering, Rensselaer Polytechnic Institute, Troy, NY 12180, USA}

\author{Andrew N. Cleland}
\email{anc@uchicago.edu}
\affiliation{Pritzker School of Molecular Engineering, University of Chicago, Chicago, IL 60637, USA}

\date{\today}


\keywords{Superconducting qubit, entanglement, multi-qubit gates}

\maketitle
\clearpage

\section{Device characteristics}
Our device schematic and wiring diagram are as shown in Refs.~\cite{Wu2024, Wu2025}, with qubits and switches tuned by external flux lines, with additional RF digital-analog converters (DACs) for driving the qubits. The $14$-bit DAC we use provides sampling rate up to about $1~\mathrm{GSPS}$, while the analog output signal passes through an additional custom designed $250~\mathrm{MHz}$ Gaussian filter. To tackle flux pulse distortion from experimental wirings and filters, we implement a pre-distortion technique~\cite{Chen2018metrology} to facilitate the deployment and benchmarking of quantum gates. Specifically, we apply a square flux pulse with a prescribed amplitude and duration to a qubit initialized on the equatorial plane of the Bloch sphere. After an idle interval, quantum state tomography is performed to reconstruct the qubit state. By analyzing the resulting state evolution, we quantify the residual flux present between the square pulse and the tomography operation, thereby enabling calibration of the distortion transfer function. Given that the flux lines share an identical wiring configuration, the level of distortion is expected to be comparable across all components. Any additional imperfections can in principle be handled by the reinforcement learning-assisted pulse shaping technique, discussed in detail later.

With the router switches turned off by applying magnetic fluxes to set each switch frequency to its (low) idle frequency, we perform single-qubit tune-up; the single-qubit metrics for all the qubits are shown in Table~\ref{tab:qubit_params}. Two-qubit operations are characterized by simultaneous fast flux pulses to the corresponding switches, turning on the interaction, with the qubits tuned to pre-defined operating frequencies. This is done in a pair-wise fashion, with minimum crosstalk to the other (idle) qubits~\cite{Wu2024}.

\begin{figure}[tb]
    \centering
    \includegraphics{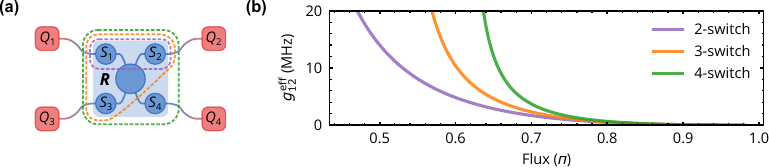}
    \caption{Qubit-qubit coupling strengths. (a) Schematic 4-qubit router; colored dashed lines enclose the switches involved in the corresponding simulations of effective $Q_1$-$Q_2$ coupling strength $g_{12}^{\mathrm{eff}}$ in (b). The coupling strength is calculated via a Schrieffer-Wolff transformation~\cite{Wu2024}.} \label{fig:fig_g_num_switch}
\end{figure}

Multi-qubit operations, especially when certain resonance conditions or coupling strength configurations must be satisfied, require more careful and complicated tune-up procedures. As all the switches hybridize strongly with one another, tuning one additional switch will typically affect the other switches. Also, as more switches are turned on, the flux pulse amplitudes required to maintain the same qubit-qubit coupling become smaller, due to the increase in frequency of the hybridized switch modes, as shown in Fig.~\ref{fig:fig_g_num_switch}.

\begin{table}[tb]
 \caption{Qubit parameters. $\omega_{q}$ ($\eta$) is the qubit idle frequency (anharmonicity), $T_{1}$ ($T_{2}^{*}$) is the qubit energy relaxation time (Ramsey decoherence time), and $\mathcal{F}_{\mathrm{SQG}}=\left(\mathcal{F}_{X}+\mathcal{F}_{X/2}\right)/2$ is the average single-qubit gate fidelity measured by randomized benchmarking. In terms of qubit readout, $\omega_{rr}$ ($\tau_{rr}$) is the readout frequency (duration), $\chi$ is the dispersive shift, $F_{g}$ ($F_{e}$) is the readout fidelity of the ground state $\ket{g}$ (excited state $\ket{e}$), defined as the measured $\ket{g}$ ($\ket{e}$) probability when the qubit is initialized to the respective state. \label{tab:qubit_params}}
 \centering
 \begin{ruledtabular}
 \bgroup
 \def\arraystretch{1.5}
 \begin{tabular}{ccccc}
   &$Q_{1}$&$Q_{2}$&$Q_{3}$&$Q_{4}$\\
   \hline
   $\omega_{rr}/2\pi$&5.7630 GHz&5.6997 GHz&5.6245 GHz&5.6753 GHz\\
   $\omega_{q}/2\pi$&4.5852 GHz&4.5409 GHz&4.7435 GHz&4.6553 GHz\\
   $\eta/2\pi$&-170 MHz&-181 MHz&-178 MHz&-163 MHz\\
   $2\chi/2\pi$&2.1 MHz&2.1 MHz&2.1 MHz&1.3 MHz\\
   $\tau_{rr}$&700 ns&920 ns&880 ns&840 ns\\
   $F_{g}$&99.7\%&99.6\%&95.7\%&98.8\%\\
   $F_{e}$&97.4\%&98.2\%&93.1\%&95.9\%\\
   $T_{1}$&36.0 $\mu$s&41.4 $\mu$s&22.1 $\mu$s&11.6 $\mu$s\\
   $T_{2}^{*}$&671 ns&965 ns&1176 ns&505 ns\\
    $\mathcal{F}_{\mathrm{SQG}}$&$99.52\%\pm0.05\%$&$99.71\%\pm0.10\%$&$99.80\%\pm0.03\%$&$99.58\%\pm0.15\%$
 \end{tabular}
 \egroup
 \end{ruledtabular}
\end{table}

To measure the effective qubit-qubit coupling strengths when more than two qubits are involved, we benchmark in a sequential order: We turn on all the relevant switches, then to measure $g_{ij}$ between $Q_i$ and $Q_j$, flux-bias the frequencies of the uninvolved qubits $Q_{k\neq i,j}$ away from $Q_i$ and $Q_j$, then observe Rabi oscillations between $Q_i$ and $Q_j$. This is repeated for all qubit pairs. As the number of qubits in the processor increases, we can use more efficient measurement schemes, such as those for Hamiltonian learning~\cite{Hangleiter2024}.

\section{Reinforcement learning setup}
Reinforcement learning (RL) provides a computational framework for sequential decision-making, where an agent interacts with an environment to learn, through trial and error, what actions will maximize a cumulative reward. RL has been increasingly applied to the interactive control tasks found in robotics, game playing, high-frequency trading, and automated driving, among others. RL can be formulated using two spaces, a state space $\mathcal{S}=\{s\}$, and an action space $\mathcal{A}=\{a\}$~\cite{Sutton2018}. The environment, characterized by a transition probability $\mathcal{P}=\mathrm{Pr}\left( s_{t+1},r_{t+1}|s_t,a_t \right)$, determines how the state $s_t$ is updated and how much reward $r_{t+1}$ the agent gains upon taking the action $a_t$. Here the subscript $t$ represents the time step. The agent's task is to learn a policy $\pi_{\theta}\left( a_t|s_t \right)$, which gives the probability the agent should take the action $a_t$ given the state $s_t$, with policy parameters $\theta$; the policy is updated as the agent learns. The ultimate goal of the learning optimization is to maximize the discounted cumulative reward $R=\sum_{t=0}^{\infty}\gamma^{t}r_{t+1}$ over all possible initial states $s_0$, where $\gamma$ is the discount factor, used to time-discount rewards obtained in the future. There are many algorithms to optimize the policy parameters $\theta$, such as Q-Learning~\cite{Watkins1992}, a value-based algorithm which learns the action-value function $Q(s,a)$ that estimates the expected return of taking action $a$ in state $s$; and policy-based algorithms such as REINFORCE~\cite{Williams1992}, actor-critic (A2C)~\cite{Konda2000}, and proximal policy optimization (PPO)~\cite{Schulman2017}, that directly optimize the policies based on the sampled trajectories. As introduced in the main text, in our experiment, the actions are the flux pulse segments applied to the quantum processor, which are continuous and thus more suitable for the implementation of policy-based algorithms. In this work we used PPO, which has previously been applied to various quantum experiments~\cite{Sivak2022, Sivak2023, Ding2023}, due to its robustness and stability in relatively noisy environments.

\subsection{Proximal Policy Optimization (PPO)}
PPO is an extension to the A2C algorithm, where two separate networks, namely the actor $\pi_{\bm{\theta}}\left( \bm{a}|\bm{s} \right)$ and the critic $V_{\bm{\phi}}\left( \bm{s} \right)$, are jointly optimized and converge together to yield a learned policy. During the process, the actor proposes actions $\bm{a}$ given states $\bm{s}$, and the critic evaluates how good those actions are based on the estimated value function $V_{\bm{\phi}}$; note we use bold fonts to indicate vectors.

As policy-based RL typically requires an on-policy training, where sampled trajectories are directly from the policy but not from other distributions, we prepare a batch of samples from the current policy for training in each epoch. In A2C, the samples are obtained by following the trajectory $\left( \bm{s}_{t}, \bm{a}_{t} \right)\rightarrow\left( \bm{s}_{t+1}, r_{t+1} \right)$ using the current policy $\pi_{\bm{\theta}}\left( \bm{a}_{t}|\bm{s}_{t} \right)$, with samples labeled as $\left( \bm{s}_{i,t_i}, \bm{a}_{i,t_i}; \bm{s}_{i,t_i+1}, r_{i,t_i+1} \right)$, $i$ representing the sample index in the batch. For each sample we define an advantage function
\begin{equation}\label{eq:advantage_func}
    A\left( \bm{s}_{i,t_i}, \bm{a}_{i,t_i} \right) = r_{i,t_i+1} + \gamma V_{\bm{\phi}}\left( \bm{s}_{i,t_i+1} \right) - V_{\bm{\phi}}\left( \bm{s}_{i,t_i} \right),
\end{equation}
which measures how much better (or worse) the action $\bm{a}_{i,t_i}$ is compared to the average action for the state $\bm{s}_{i,t_i}$. The critic network is then updated by minimizing the objective function
\begin{equation}\label{eq:objective_func}
    \mathcal{L}(\bm{\phi}) = \frac{1}{2N_b}\sum_{i}A^{2}\left( \bm{s}_{i,t_i}, \bm{a}_{i,t_i} \right),
\end{equation}
where $N_b$ is the batch size. To optimize the actor network (the policy), we define a utility function
\begin{equation}\label{eq:utility_func_A2C}
    J_{\mathrm{A2C}}(\bm{\theta}) = \frac{1}{N_b}\sum_i A\left( \bm{s}_{i,t_i}, \bm{a}_{i,t_i} \right) \ln \pi_{\bm{\theta}}\left( \bm{a}_{i,t_i} | \bm{s}_{i, t_i} \right),
\end{equation}
and then apply gradient ascent to $J_{\mathrm{A2C}}(\bm{\theta})$ to update $\bm{\theta}$.

We note from Eq.~(\ref{eq:utility_func_A2C}) that the utility function is evaluated using the same policy to generate the samples, which is considered to be less robust when interacting with a noisy environment. PPO therefore introduces an additional policy and a clip function to tackle this. In PPO, in addition to the current policy $\pi_{\bm{\theta}}\left( \bm{a}_{t}|\bm{s}_{t} \right)$ that will be updated, we have an old policy $\pi_{\bm{\theta}}^{\mathrm{old}}\left( \bm{a}_{t}|\bm{s}_{t} \right)$ which is used to generate the samples in the batch. The old policy is set to be equal to the current policy at the beginning of the training, and the old policy updated with the current policy every $K$ epochs. 

The clipped PPO utility function is given by
\begin{equation}\label{eq:utility_func_PPO}
    J_{\mathrm{PPO}}(\bm{\theta}) = \frac{1}{N_b}\sum_{i}\mathrm{min}\left[ r_{i}(\bm{\theta})A(\bm{s}_{i,t_i}, \bm{a}_{i,t_i}), \mathrm{clip}\left( r_{i}(\bm{\theta}), 1-\epsilon, 1+\epsilon \right)A(\bm{s}_{i,t_i}, \bm{a}_{i,t_i}) \right],
\end{equation}
where $r_{i}(\bm{\theta})=\pi_{\bm{\theta}}\left( \bm{a}_{i,t_i}|\bm{s}_{i,t_i} \right)/\pi_{\bm{\theta}}^{\mathrm{old}}\left( \bm{a}_{i,t_i}|\bm{s}_{i,t_i} \right)$ is the probability ratio and $\epsilon$ is the clipping factor. Similarly, $\bm{\theta}$ is updated by applying gradient ascent to $J_{\mathrm{PPO}}(\bm{\theta})$. With the clipped utility function, large policy updates are suppressed, yielding better policy stability in the presence of environmental noise. Note that in the actual experiments, we instead apply gradient descent to $-J_{\mathrm{PPO}}(\bm{\theta})$.

\begin{algorithm}
\caption{Optimization process with PPO}
\label{alg:RL}
\KwIn{Initial policy $\bm{a}_{\mathrm{mean}}(0)$ and $\bm{a}_{\mathrm{std}}(0)$; Initial value $\mathcal{V}(0)$; Learning rates $\eta_a,\eta_c$; PPO parameters $K,\epsilon,N_b$}
\KwOut{Optimized control policy $\bm{a}_{\mathrm{mean}}^{*}$}
Initialize epoch $t \gets 1$; current policy $\bm{a}_{\mathrm{mean}}(t) \gets \bm{a}_{\mathrm{mean}}(0)$ and $\bm{a}_{\mathrm{std}}(t) \gets \bm{a}_{\mathrm{std}}(0)$; old policy $\bm{a}_{\mathrm{mean}}^{\mathrm{old}} \gets \bm{a}_{\mathrm{mean}}(t)$ and $\bm{a}_{\mathrm{std}}^{\mathrm{old}} \gets \bm{a}_{\mathrm{std}}(t)$; value $\mathcal{V}(t) \gets \mathcal{V}(0)$\;
\For{$t \gets 1$ \KwTo $T-1$}{
   Sample $N_b$ actions $\bm{a}(t;i=1:N_b)$ from $\bm{a}_{\mathrm{mean}}^{\mathrm{old}}$ and $\bm{a}_{\mathrm{std}}^{\mathrm{old}}$\;
   Calculate the corresponding probability ratios with $\bm{a}_{\mathrm{mean}}(t),\bm{a}_{\mathrm{std}}(t),\bm{a}_{\mathrm{mean}}^{\mathrm{old}},\bm{a}_{\mathrm{std}}^{\mathrm{old}}$\;
   Obtain $N_b$ rewards $r(t;i=1:N_b)$ by evolving the quantum system with corresponding actions $\bm{a}(t;i=1:N_b)$\;
   \eIf{$K|t$}{
    $\bm{a}_{\mathrm{mean}}^{\mathrm{old}} \gets \bm{a}_{\mathrm{mean}}(t)$ and $\bm{a}_{\mathrm{std}}^{\mathrm{old}} \gets \bm{a}_{\mathrm{std}}(t)$\;
    }{
        Pass\;
    }
    Obtain gradients $\delta\bm{a}_{\mathrm{mean}},\delta\bm{a}_{\mathrm{std}},\delta\mathcal{V}$ from Eqs.~(\ref{eq:objective_func}) and (\ref{eq:utility_func_PPO})\;
    $\bm{a}_{\mathrm{mean}}(t+1) \gets \bm{a}_{\mathrm{mean}}(t) - \eta_{a}\delta\bm{a}_{\mathrm{mean}}$; $\bm{a}_{\mathrm{std}}(t+1) \gets \bm{a}_{\mathrm{std}}(t) - \eta_{a}\delta\bm{a}_{\mathrm{std}}$; $\mathcal{V}(t+1) \gets \mathcal{V}(t) - \eta_{c}\delta\mathcal{V}$\footnote{Here we use the ADAM optimizer~\cite{Kingma2017} for more efficient updates}\;
    $t \gets t+1$\;
}
\Return $\bm{a}_{\mathrm{mean}}(T)$\;
\end{algorithm}

\subsection{Training process flow}
In the experiment, we focus on quantum operations that conserve  excitation number in the qubit subspace, so can be realized solely by manipulating the fluxes in the SQUID loops of the qubits and switches, with actions parameterized with piecewise-constant pulse segments. We design the RL process as a stateless approach, which means instead of a multi-stage optimization where the action is chosen according to the current state of the system, we directly optimize the action that governs the quantum dynamics. Specifically, as illustrated in Fig. 3 of the main text, instead of using a neural network to formulate the policy $\pi_{\bm{\theta}}(\bm{a}|\bm{s})$, we use as training parameters two vectors $\bm{a}_{\mathrm{mean}}$ and $\bm{a}_{\mathrm{std}}$, which have the same dimensions as the flux pulse segments. Similarly, for the value function $V_{\bm{\phi}}(\bm{s})$, we simplify by training a single value $\mathcal{V}$, which is equivalently a direct measure of the operation fidelity. In Algorithm~\ref{alg:RL} we summarize the corresponding pseudocode of the optimization process. In the experiment, we set $\eta_a=0.0001, \eta_c=0.001, \epsilon=0.2, K=1, N_b=10$, and we decrease the learning rates over time to control the stability of the training process. For most of the tasks, we assign eight segments to each pulse (including two zero paddings at the beginning and the end). If the pulse is $50$-$\mathrm{ns}$ long, major value changes can occur every $50/(8-1)\approx7~\mathrm{ns}$. This corresponds to at most a $1/(2\times7)~\mathrm{ns}\approx70~\mathrm{MHz}$ pulse bandwidth, which is generally safe even for shorter pulses, considering the $250~\mathrm{MHz}$ cutoff frequency of the Gaussian filter.

\section{GHZ state optimization}
\begin{figure}[tb]
    \centering
    \includegraphics{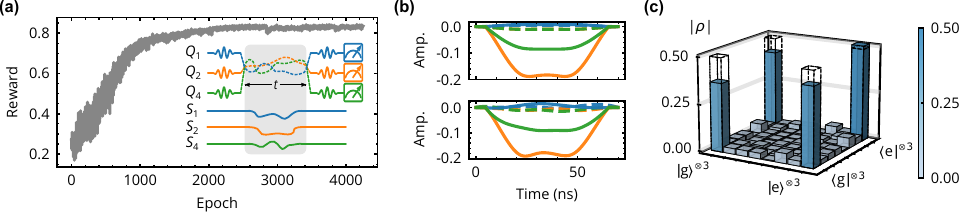}
    \caption{Three-qubit GHZ state generation with RL. (a) Reward evolution; inset shows pulse sequences. Shaded area shows optimized portion of pulse segments. (b) Pulses before (top) and after (bottom) optimization. (c) Tomography of the optimized GHZ state, with a fidelity of $\mathcal{F}=\mathrm{tr}\left( \rho_{\mathrm{ideal}}\rho \right)=86.58(8)\%$.} \label{fig:fig_GHZ}
\end{figure}

Our one-step GHZ state preparation is inspired by Ref.~\cite{Song2019}, where the one-axis twisting Hamiltonian is synthesized using all-to-all transverse couplings. In that demonstration, the qubits are initially prepared on the Bloch sphere equator with equal phases, then brought on resonance with equal mutual coupling strengths $g$ for a duration $\pi/2g$. Finally, $\pi/2$ pulses with phases applied to the qubits depending on the qubit number to realize a GHZ state. As mentioned above, due to the requirement that qubits are both on-resonance and have equal coupling at the same time, this protocol becomes more challenging when many switches are involved. We therefore deploy the RL method described in the previous section. We fix the phases for the initial and final $\pi/2$ pulses, and optimize the flux pulses in between, as shown inset in Fig.~\ref{fig:fig_GHZ}(a). We parameterize each flux pulse with a piecewise constant function of $6$ segments, then smooth by interpolation. This means $\bm{a}_{\mathrm{mean}}$ and $\bm{a}_{\mathrm{std}}$ will each consist of $6\times6=36$ elements, as there are $6$ pulses in total. We define the reward as $r=1-\abs{P(\ket{000})-0.5}-\abs{P(\ket{111})-0.5}$. We then assign an initial seed (upper part of Fig.~\ref{fig:fig_GHZ}(b)) to the agent and train the state preparation according to Algorithm~\ref{alg:RL}. The reward evolution is shown in Fig.~\ref{fig:fig_GHZ}(a), which saturates after about $3000$ epochs. The resulting optimized pulses are shown in the lower part of Fig.~\ref{fig:fig_GHZ}(b), with state tomography in Fig.~\ref{fig:fig_GHZ}(c). The benchmarked state fidelity is $\mathcal{F}=\mathrm{tr}\left( \rho_{\mathrm{ideal}}\rho \right)=86.58(8)\%$. Notably, the prepared state has a higher probability to be in $\ket{111}$ compared to $\ket{000}$, which is possibly due to overcorrection of the otherwise naturally lower $P(\ket{111})$ due to finite qubit $T_1$, where the agent tries to increase this population while lowering $P(\ket{000})$ slightly. This can in principle be improved by assigning different weights to the terms in the reward function involving $P(\ket{000})$ and $P(\ket{111})$.

\begin{figure}[tb]
    \centering
    \includegraphics{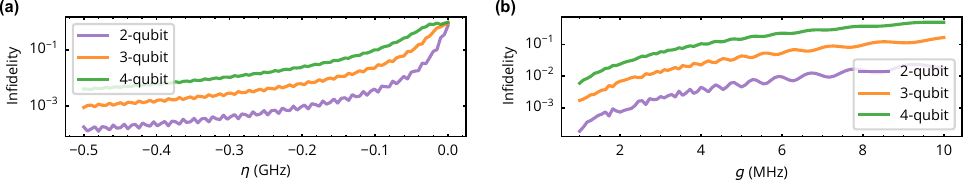}
    \caption{Simulation of GHZ state generation with the one-axis-twisting Hamiltonian synthesis protocol~\cite{Song2019}. (a) Infidelity as a function of qubit anharmonicity $\eta$, where the coupling strength $g$ is chosen to be $2~\mathrm{MHz}$. (b) Infidelity as a function of $g$, where $\eta=-0.2~\mathrm{GHz}$.} \label{fig:fig_GHZ_fidsimu}
\end{figure}

We note that the anharmonicity $\eta$ of the qubits and their coupling strength $g$ (assuming uniformity) can strongly affect the state preparation fidelity. We simulate an $N$-qubit all-to-all coupled system using the rotating-frame Hamiltonian
\begin{equation}
    \mathcal{H} = \sum_i \frac{\eta}{2}a_i^{\dagger}a_i^{\dagger}a_ia_i + g\sum_{i<j}\left( a_i^{\dagger}a_j + a_j^{\dagger} a_i\right),
\end{equation}
where the qubits are assumed to be on-resonance during the interaction. The initial and final $\pi/2$ pulses are applied accordingly. No decoherence is considered and the fidelity is measured by $\abs{\braket{\varphi_{\mathrm{final}}}{\varphi_{\mathrm{ideal}}}}^2$. The results are plotted in Fig.~\ref{fig:fig_GHZ_fidsimu}, where we see that in general fidelity increases with larger $\eta$ and smaller $g$. However, in our setup, $\eta$ is fixed (see Table~\ref{tab:qubit_params}) and smaller $g$ is not ideal because this results in a longer preparation time; this, together with the limited qubit coherence, reduces the final state fidelity.

\section{XEB and OXEBIT}
In this section, we describe two methods we used for the optimization of entangling gates: Cross-entropy benchmarking (XEB)~\cite{Boixo2018, Arute2019} and optimized cross-entropy benchmarking for immediate tuneup (OXEBIT), following the spirit of optimized randomized benchmarking for immediate tuneup (ORBIT)~\cite{Kelly2014}.

\subsection{Cross-entropy benchmarking (XEB)}
XEB was first introduced to benchmark the dynamics of a large quantum processor, but given its flexibility, it is also widely used for benchmarking quantum gate fidelities, especially those outside the Clifford group. A general procedure for XEB can be found in Appendix F of Ref.~\cite{Wu2024}. The fidelity of a quantum gate $\mathcal{U}$ can be calculated from a comparison of the reference and interleaved benchmarking decay rates, given by
\begin{equation}\label{eq:XEB_gate_fid}
    \mathcal{F}(\mathcal{U}) = 1 - \frac{d-1}{d}\left( 1 - \frac{p_{\mathcal{U}}}{p_{\mathrm{ref}}} \right),
\end{equation},
where $d$ is the Hilbert space dimension, and $p_{\mathcal{U}}$ ($p_{\mathrm{ref}}$) is the interleaved (reference) sequence decay constant, extracted from an exponential fit to the sequence fidelity
\begin{equation}\label{eq:XEB_p}
    \mathcal{F}(m) = A \cdot p^m + B,
\end{equation}
where $m$ is the circuit depth, and $A$ and $B$ are fitting parameters related to state preparation and measurement (SPAM) errors. We estimate the uncertainty of the gate fidelity by adapting the \emph{bootstrapping} method~\cite{Barends2014}. Specifically, in our experiment, for both $p_{\mathcal{U}}$ and $p_{\mathrm{ref}}$, we resample $1000$ datasets with replacement from the respective original dataset, from which, using Eq.~(\ref{eq:XEB_p}), we extract an array of $1000$ $p_{\mathcal{U}}$ and $p_{\mathrm{ref}} values$. We then calculate the mean values $\mu(p_{\mathcal{U}}), \mu(p_{\mathrm{ref}})$ and the standard deviation $\sigma(p_{\mathcal{U}}), \sigma(p_{\mathrm{ref}})$, from which we can compute the mean and standard deviation of $\mathcal{F}(\mathcal{U})$ according to Eq.~(\ref{eq:XEB_gate_fid}), given by
\begin{equation}\label{eq:XEB_gate_fid_mean_std}
    \begin{aligned}
        \mu\left( \mathcal{F}(\mathcal{U}) \right) &= 1 - \frac{d-1}{d}\left( 1 - \frac{\mu(p_{\mathcal{U}})}{\mu(p_{\mathrm{ref}})} \right),\\
        \sigma\left( \mathcal{F}(\mathcal{U}) \right) &= \frac{d-1}{d}\sqrt{\frac{\mu^2(p_{\mathcal{U}})}{\mu^2(p_{\mathrm{ref}})}\left( \frac{\sigma^2(p_{\mathcal{U}})}{\mu^2(p_{\mathcal{U}})} + \frac{\sigma^2(p_{\mathrm{ref}})}{\mu^2(p_{\mathrm{ref}})} \right)}.
    \end{aligned}
\end{equation}

\subsection{Optimized cross-entropy benchmarking for immediate tuneup (OXEBIT)}
The OXEBIT method is directly adapted from ORBIT, an RB-based closed-loop optimization method for quantum gates, proposed and demonstrated in Ref.~\cite{Kelly2014}. The core idea is to use the sequence fidelity in Eq.~(\ref{eq:XEB_p}) at a given circuit depth $m$ as a proxy for the gate fidelity. Since this process is fast and SPAM-free, it enables more efficient optimization of the gate fidelity while being compatible with a variety of optimization methods.

The detailed process in each epoch is as follows:
\begin{enumerate}
    \item Select a batch of pulse shapes from the mean and standard deviation values of the pulse segments.
    \item Generate random numbers for XEB with certain $m$ and $k$ values.
    \item For each sample in the batch, execute the XEB pulse sequence and obtain the reward defined in the main text. Note that this reward definition is proportional but not equal to the sequence fidelity, and they coincide at value $1$.
    \item Update training parameters using the batch of rewards and previously generated pulse shapes.
\end{enumerate}
It is worth mentioning that during the optimization process, we use a reduced number of random circuits per epoch ($k=10$) compared to a full benchmarking ($k=100$) to accelerate the training cycle while maintaining a reasonable signal-to-noise ratio. Owing to the complexity of multi-qubit entangling operations, we regenerate fresh random numbers at each epoch to ensure the model continues to converge toward the target quantum gate. For each data point, we execute the circuit $1020$ times to get the statistics.

\section{CZ gate optimization}

\begin{figure}[tb]
    \centering
    \includegraphics{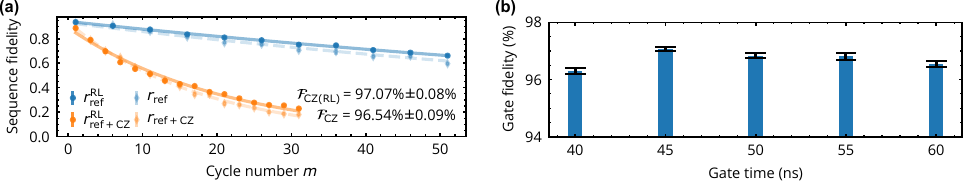}
    \caption{CZ gate optimization with qubits $Q_1$-$Q_2$. (a) XEB data with both RL-optimized and hand-tuned gates. (b) RL-optimized gate fidelity as a function of the gate time.} \label{fig:fig_CZ}
\end{figure}

In Fig.~\ref{fig:fig_CZ}(a) we display the results for RL optimization of the CZ gate between $Q_1$ and $Q_2$. Using the hand-tuned parameters as a guide for initializing the training parameters, we can achieve higher gate fidelity with less human intervention. Moreover, we vary the gate time and repeat the training process with the same initial parameters, where we find an apparent tradeoff between gate time and fidelity, likely balancing qubit coherence (shorter gate times) with population leakage into the router (longer gate times). The relevant results are shown in Fig.~\ref{fig:fig_CZ}(b).

\section{Three-qubit gate optimization}
\subsection{CSWAP}
The controlled-SWAP (CSWAP) gate can play a pivotal role in the development of bucket-brigade quantum random access memory (QRAM)~\cite{Giovannetti2008}, where the control qubit, set with the memory address, controls how the data qubit is routed, using a bifurcated tree structure. In our setup, as mentioned in the main text, the CSWAP gate is optimized using different initial computational basis states. Here, we show how we determine the various phases accumulated during the process. There are a total of seven phases: $\phi_{001}, \phi_{010}, \phi_{011}, \dots, \phi_{111}$. In addition, using the rotating frame results in phase accumulation~\cite{Wu2024} for transversal gates like CSWAP when the idle frequencies of the qubits differ. Altogether, the matrix representation of the CSWAP gate in the computational basis is given by
\begin{equation}\label{eq:U_CSWAP}
    \mathcal{U}_{\mathrm{CSWAP}} = 
    \begin{pmatrix}
        1 & 0 & 0 & 0 & 0 & 0 & 0 & 0\\
        0 & e^{i\phi_{001}} & 0 & 0 & 0 & 0 & 0 & 0\\
        0 & 0 & e^{i\phi_{010}} & 0 & 0 & 0 & 0 & 0\\
        0 & 0 & 0 & e^{i\phi_{011}} & 0 & 0 & 0 & 0\\
        0 & 0 & 0 & 0 & e^{i\phi_{100}} & 0 & 0 & 0\\
        0 & 0 & 0 & 0 & 0 & 0 & e^{i(\phi_{110} - \Delta_{23}t_b)} & 0\\
        0 & 0 & 0 & 0 & 0 & e^{i(\phi_{101} + \Delta_{23}t_b)} & 0 & 0\\
        0 & 0 & 0 & 0 & 0 & 0 & 0 & e^{i\phi_{111}}
    \end{pmatrix},
\end{equation}
where $\Delta_{23}$ is the difference between the idle frequencies of the second and the third qubits, and $t_b$ is the start time of the gate. The procedure to determine the various phases is as follows:
\begin{enumerate}
    \item For $\phi_{001}$, we prepare $\ket{\varphi} = \left( \ket{000} - i\ket{001} \right) / \sqrt{2}$ by applying an $X/2$ pulse to the third qubit; then apply the CSWAP gate, after which the quantum state becomes $\ket{\varphi} = \left( \ket{000} - ie^{i\phi_{001}}\ket{001} \right) / \sqrt{2}$; finally we apply another $X(\phi)/2$ pulse ($\phi$ is the phase shift from the $X$ axis), and try to maximize $P(\ket{1})$ for the third qubit by sweeping $\phi$; the optimal value is $\phi_{001}$.
    \item For $\phi_{010}$, we prepare $\ket{\varphi} = \left( \ket{000} - i\ket{010} \right) / \sqrt{2}$ by applying an $X/2$ pulse to the second qubit; we then apply the CSWAP gate, after which the quantum state becomes $\ket{\varphi} = \left( \ket{000} - ie^{i\phi_{010}}\ket{010} \right) / \sqrt{2}$; finally we apply another $X(\phi)/2$ pulse and try to maximize $P(\ket{1})$ for the second qubit by sweeping $\phi$, where the optimal value is $\phi_{010}$.
    \item For $\phi_{100}$, we prepare $\ket{\varphi} = \left( \ket{000} - i\ket{100} \right) / \sqrt{2}$ by applying an $X/2$ pulse to the first qubit; we then apply the CSWAP gate, after which the quantum state becomes $\ket{\varphi} = \left( \ket{000} - ie^{i\phi_{100}}\ket{100} \right) / \sqrt{2}$; finally we apply another $X(\phi)/2$ pulse and try to maximize $P(\ket{1})$ for the first qubit by sweeping $\phi$, where the optimal value is $\phi_{100}$.
    \item For $\phi_{011}$, we prepare $\ket{\varphi} = \left( \ket{001} - i\ket{011} \right) / \sqrt{2}$ by applying an $X/2$ pulse to the second qubit and an $X$ pulse to the third qubit; we then apply the CSWAP gate, after which the quantum state becomes $\ket{\varphi} = \left( \ket{001} - ie^{i(\phi_{011} - \phi_{001})}\ket{011} \right) / \sqrt{2}$; finally we apply another $X(\phi)/2$ pulse and try to maximize $P(\ket{1})$ for the second qubit by sweeping $\phi$, where the optimal value is $\phi_{011} - \phi_{001}$.
    \item For $\phi_{111}$, we prepare $\ket{\varphi} = \left( \ket{011} - i\ket{111} \right) / \sqrt{2}$ by applying an $X/2$ pulse to the first qubit and an $X$ pulse to both the second and the third qubit; we then apply the CSWAP gate, after which the quantum state becomes $\ket{\varphi} = \left( \ket{011} - ie^{i(\phi_{111} - \phi_{011})}\ket{111} \right) / \sqrt{2}$; finally we apply another $X(\phi)/2$ pulse and try to maximize $P(\ket{1})$ for the first qubit by sweeping $\phi$, with optimal value $\phi_{111} - \phi_{011}$.
    \item For $\phi_{101}$, we prepare $\ket{\varphi} = \left( \ket{100} - i\ket{101} \right) / \sqrt{2}$ by applying an $X/2$ pulse to the third qubit and an $X$ pulse to the first qubit; we then apply the CSWAP gate, after which the quantum state becomes $\ket{\varphi} = \left( \ket{100} - ie^{i(\phi_{101} - \phi_{100} + \Delta_{23}t_b)}\ket{110} \right) / \sqrt{2}$; finally we apply another $X(\phi + \Delta_{23}t_b)/2$ pulse and try to maximize $P(\ket{1})$ for the second qubit by sweeping $\phi$, with optimal value $\phi_{101} - \phi_{100}$.
    \item For $\phi_{110}$, we prepare $\ket{\varphi} = \left( \ket{100} - i\ket{110} \right) / \sqrt{2}$ by applying an $X/2$ pulse to the second qubit and an $X$ pulse to the first qubit; we then apply the CSWAP gate, after which the quantum state becomes $\ket{\varphi} = \left( \ket{100} - ie^{i(\phi_{110} - \phi_{100} - \Delta_{23}t_b)}\ket{101} \right) / \sqrt{2}$; finally we apply another $X(\phi - \Delta_{23}t_b)/2$ pulse and try to maximize $P(\ket{1})$ for the third qubit by sweeping $\phi$, with optimal value $\phi_{110} - \phi_{100}$.
\end{enumerate}
Once all the phases are determined, we use Eq.~(\ref{eq:U_CSWAP}) as the ideal unitary matrix to benchmark the gate fidelity with XEB. To estimate the decoherence impact on the gate fidelity, we use Eq.~(12) in Ref.~\cite{Abad2022}, which reads $\mathcal{F}_{\mathrm{dec}}=1 - \frac{d}{2\left(d+1\right)}\tau\sum_{k=1}^{N}\left( \Gamma_1^k + \Gamma_{\phi}^k \right)$, where $N$ is the qubit number, $d=2^N$ is the system dimension, $\tau$ is the gate duration, and $\Gamma_1^k$ ($\Gamma_{\phi}^k$) is the energy relaxation (pure dephasing) rate of the $k$th qubit. Plugging in the qubit parameters from Table~\ref{tab:qubit_params} and $\tau=40~\mathrm{ns}$, we have $1-\mathcal{F}_{\mathrm{dec}}\approx8.14\%$, which shows a decoherence-dominated gate fidelity. We have considered providing a more comprehensive evaluation of the error budget, especially for the three-qubit gates, via more quantitative numerical simulations. However, due to the non-intuitive RL-optimized pulses and incomplete knowledge of circuit parameters, this is difficult to do unambiguously. Furthermore, simulating leakage outside the qubit computational basis and into the switch modes would involve at least six coupled three-level systems, which is computationally hard. Finally, we find that the qubit dephasing depends strongly on how close we tune the switch modes to the qubit modes, which is nontrivial to capture in simulations.

\subsection{CCPHASE}
\begin{figure}[tb]
    \centering
    \includegraphics{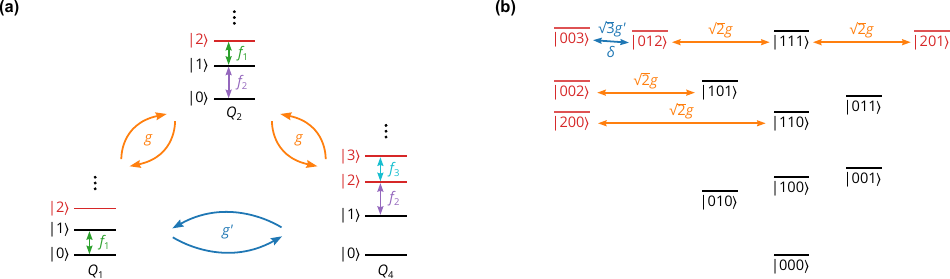}
    \caption{CCPHASE gate schematic. (a) Qubit frequency configuration for the initial seed of RL. Transitions with the same frequencies are labeled with the same colors. (b) Energy diagram of the three-qubit system, labeled as $\ket{Q_2Q_1Q_4}$. Black (red) states indicate the (non-)computational basis states.} \label{fig:fig_CCPHASE_schematic}
\end{figure}

The CCPHASE gate is the building block for the Toffoli (CCNOT) gate, which is essential in e.g. quantum error correction as a syndrome extraction tool. In our experiment, we use $Q_1$ and $Q_2$ as the control qubits and $Q_4$ as the target qubit, where we write the system state as $\ket{Q_2Q_1Q_4}$. As mentioned in the main text, the CCPHASE gate we optimize can be written as $\mathcal{U}_{\mathrm{CCPHASE}}=\dyad{10}{10}\otimes\left( -\sigma_z \right) + \left( \mathds{1} - \dyad{10}{10} \right)\otimes\mathds{1}$, where we only accumulate a $\pi$ phase when the three-qubit state is $\ket{100}$. This is inspired by the specific frequency arrangement used as an initial seed for the RL optimization process, depicted in Fig.~\ref{fig:fig_CCPHASE_schematic}(a), where we have $\omega_{01}^{Q_1}=\omega_{12}^{Q_2}$ and $\omega_{01}^{Q_2}=\omega_{12}^{Q_4}$, as well as $g_{12}=g_{24}$. The energy diagram of the system is illustrated in Fig.~\ref{fig:fig_CCPHASE_schematic}(b), where black (red) states are within (outside) the computational basis. With this frequency arrangement, $\ket{110}$ ($\ket{101}$) interacts with $\ket{200}$ ($\ket{002}$) with a coupling strength $\sqrt{2}g$, which therefore completes a round trip $\ket{110} \rightarrow -i\ket{200} \rightarrow -\ket{110}$ ($\ket{101} \rightarrow -i\ket{002} \rightarrow -\ket{101}$) in a time $\pi/\sqrt{2}g$. The three-excitation state $\ket{111}$, however, interacts with both $\ket{012}$ and $\ket{201}$, thus accelerating the effective swap rate. This is mitigated by a third transition $\ket{012}\leftrightarrow\ket{003}$ with a slight detuning $\delta=f_1-f_3$ (see Fig.~\ref{fig:fig_CCPHASE_schematic}(a)). We then optimize with RL such that $\ket{110}$, $\ket{101}$, and $\ket{111}$ complete a round trip with the same interaction time. This differs from the $\mathcal{U}_{\mathrm{CCPHASE}}$ defined above by a simple $\sigma_z$ on the first qubit, which can be easily integrated into the XEB process of the training.

By using the same method as CSWAP and $\tau=50~\mathrm{ns}$, we can estimate the decoherence contribution to the gate error as $1-\mathcal{F}_{\mathrm{dec}}\approx10.17\%$, again indicating a decoherence dominated gate fidelity.

\section{Training runtime budget}
\begin{table}[b]
 \caption{Runtime budget for various RL-optimized quantum operations. Here, time corresponds to the unit processing time per epoch per batch. \label{tab:runtime_budget}}
 \centering
 \begin{ruledtabular}
 \bgroup
 \def\arraystretch{1.5}
 \begin{tabular}{lcccc}
   &CZ&CSWAP&CCPHASE&GHZ\\
   \hline
   data acquisition (s) & $3.37\pm0.02$ & $4.58\pm0.02$ & $7.89\pm0.03$ & $0.58\pm0.01$\\
   sampling (ms) & $2.48\pm0.28$ & $3.61\pm0.34$ & $3.31\pm0.34$ & $3.55\pm0.31$\\
   RL optimization (ms) & $0.68\pm0.10$ & $0.90\pm0.09$ & $0.98\pm0.09$ & $0.94\pm0.12$\\
   data transfer (ms) & $0.21\pm0.09$ & $0.19\pm0.09$ & $0.41\pm0.16$ & $0.21\pm0.08$\\
 \end{tabular}
 \egroup
 \end{ruledtabular}
\end{table}

In the previous sections, we examined various quantum operations optimized by an RL agent. In this section, we provide a detailed breakdown of the time required for the different stages of the training process for each operation.

\begin{figure}[tb]
    \centering
    \includegraphics{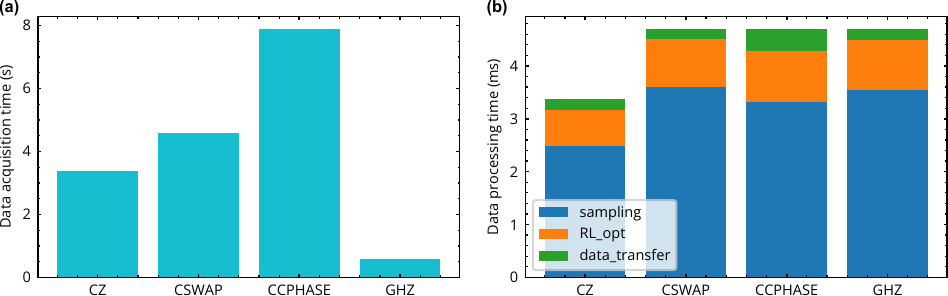}
    \caption{Runtime budget diagram. (a) Data acquisition time and (b) Data processing time, plotted with data drawn from the mean values in Table~\ref{tab:runtime_budget}.}
    \label{fig:fig_runtime_budget}
\end{figure}

The runtime budget comprises two components. The first involves data acquisition, in which pulse sequences are executed on the quantum hardware using the sampled pulse shapes, and the resulting objective function values are collected in order to update the training parameters. The second component involves data processing, which includes pre-processing (sampling pulse shapes from their probability distributions), and post-processing (transferring the acquired data to local storage), and finally updating the training parameters based on their gradients. We summarize these values in Table~\ref{tab:runtime_budget} and Fig.~\ref{fig:fig_runtime_budget}. We note that since we have a moderate number of training parameters, the majority of the optimization runtime involves data acquisition from the measurement hardware. Also, the state generation task requires less data acquisition overhead because it doesn't require scrambling of quantum states at the beginning, which is however typical for quantum gate optimization.

\section{Analog quantum dynamics}
Here we provide more details on using our setup as an analog quantum dynamics testbed, where we use the model in Ref.~\cite{Liu2020} as an example. An all-to-all connected three-qubit system with frequency modulation can be modeled using the Hamiltonian
\begin{equation}\label{eq:AQD_H1}
    \mathcal{H} = \sum_{j=1,2,3}\Delta_{j}\cos\left( \Omega_{j}t + \phi_j \right)\sigma_j^{+}\sigma_j^{-} + \sum_{jk}g_{jk}\left( \sigma_j^{+}\sigma_k^{-} + \sigma_j^{-}\sigma_k^{+} \right),
\end{equation}
where $\Delta_j$, $\Omega_j$, and $\phi_j$ are the modulation amplitude, frequency and phase, respectively. We assume that all qubits are brought on resonance during the interaction. When there is just one excitation in the system, we can consider the Hilbert space spanned by $\left\{ \ket{000}, \ket{100}, \ket{010}, \ket{001} \right\}$ ($\ket{000}$ is needed to include qubit decay). Then Eq.~(\ref{eq:AQD_H1}) can be written in matrix form as
\begin{equation}\label{eq:AQD_H2}
    \mathcal{H} =
    \begin{pmatrix}
        0 & 0 & 0 & 0\\
        0 & \Delta_1\cos\left( \Omega_1t+\phi_1 \right) & g_{12} & g_{13}\\
        0 & g_{12} & \Delta_2\cos\left( \Omega_2t+\phi_2 \right) & g_{23}\\
        0 & g_{13} & g_{23} & \Delta_3\cos\left( \Omega_3t+\phi_3 \right)
    \end{pmatrix}.
\end{equation}
We note that as we use the qubits $Q_1, Q_2$ and $Q_4$, the subscripts ``$3$'' in Eqs.~(\ref{eq:AQD_H1}) and (\ref{eq:AQD_H2}) become ``$4$'' later.

\subsection{Spin chirality}
For the implementation of spin chirality, we arrange $\Delta_j$, $\Omega_j$, and $\phi_j$ such that all the two-body interactions are eliminated. To generate the simulation for fitting the experimental data in Fig.~5 of the main text, we use $\Delta_{1,2,4}/2\pi = 140, 130, 140~\mathrm{MHz}$, $\Omega_1 = \Omega_2 = \Omega_4 = 2\pi\times100~\mathrm{MHz}$, $g_{12} = g_{14} = g_{24} = 2\pi\times14~\mathrm{MHz}$, and $\phi_{1,2,4} = 2\pi/3, 0, 4\pi/3$. Decoherence is modeled using Lindblad master equation.

\subsection{Tunable interaction by frequency modulation}

\begin{figure}[tb]
    \centering
    \includegraphics{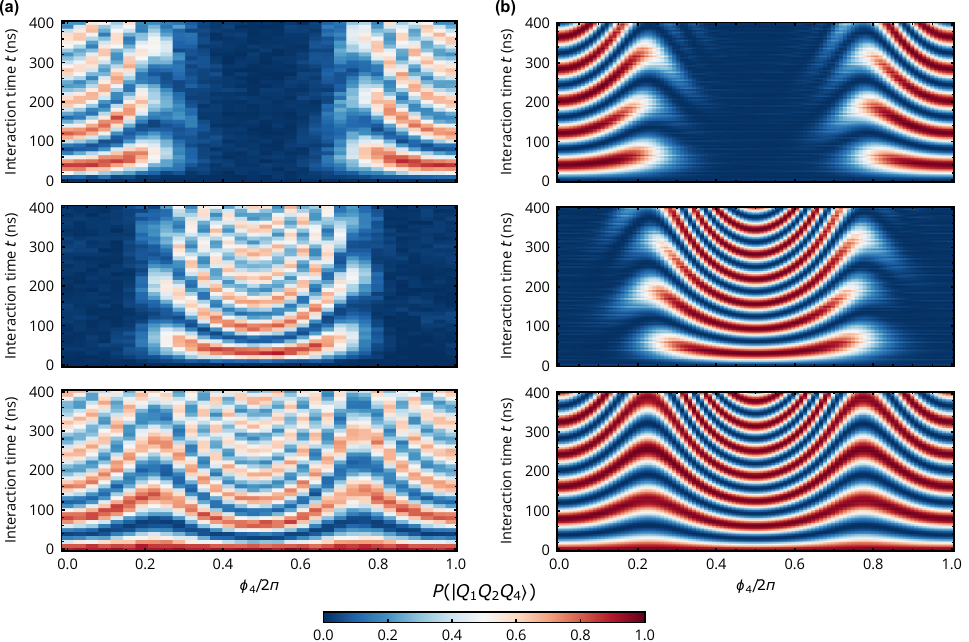}
    \caption{Tunable interaction by frequency modulation. (a) Experimental data showing the population of $P\left( \ket{Q_1Q_2Q_4} = \ket{100} \right)$ (top), $P\left( \ket{010} \right)$ (middle), and $P\left( \ket{001} \right)$ (bottom). (b) Simulation using the Hamiltonian model in Eq.~(\ref{eq:AQD_H2}).} \label{fig:fig_freqmod}
\end{figure}

In addition to three-body interactions, we can also realize a tunable two-body interaction when all three qubits are coupled simultaneously, by simply arranging the modulation phases $\phi_j$. In this case, we focus on the effective coupling strength $g_{jk}^{\mathrm{eff}}$ in the zeroth-order Hamiltonian under Floquet modulation, which reads
\begin{equation}\label{eq:AQD_geff}
    g_{jk}^{\mathrm{eff}} \simeq g_{jk}J_0\left( 2\frac{\Delta}{\Omega}\sin\frac{\phi_j - \phi_k}{2} \right),
\end{equation}
where $J_0$ is the zeroth-order Bessel function of the first kind. Here, we assume uniform $\Delta_j=\Delta$ and $\Omega_j=\Omega$. If we set $J_0(2\Delta/\Omega)=0$, $\phi_1=0$, and $\phi_2=\pi$, then by adjusting $\phi_4$, the coupling can be turned on from $Q_1$-$Q_4$ ($\phi_4=0,2\pi$) to $Q_2$-$Q_4$ ($\phi_4=\pi$). This is realized as shown in Fig.~\ref{fig:fig_freqmod}(a), where we excite $Q_4$ at the beginning, and turn on the modulation pulses to swap the excitation to either $Q_1$ or $Q_2$ depending on the value of the $Q_4$ modulation phase $\phi_4$. In Fig.~\ref{fig:fig_freqmod}(b), we simulate this effect using Eq.~(\ref{eq:AQD_H2}), with parameters $\Delta_1 = \Delta_2 = \Delta_4 = 2\pi\times120~\mathrm{MHz}$, $\Omega_1 = \Omega_2 = \Omega_4 = 2\pi\times100~\mathrm{MHz}$, $g_{12,14,24}/2\pi = 7.0, 6.2, 8.0~\mathrm{MHz}$, and $\phi_{1,2} = 0, \pi$. We do not incorporate qubit decoherence in the simulation, and extra deviations come from imperfect readout and pulse distortions. Overall, the simulated pattern shows good agreement with the experimental data.

\bibliography{ref}